\documentstyle[12pt,preprint,aps,prb]{revtex}
\def\nm{\nonumber}
\def\beqa{\begin{eqnarray}}
\def\beq{\begin{equation}}
\def\F{{\cal{F}}}      
\def\eeqa{\end{eqnarray}}
\def\eeq{\end{equation}}
\def\lab{\label}    
\def\pa{\partial}

\def\l{\Lambda}
\def\Del{\Delta}

\begin{document}

\begin{titlepage}
\thispagestyle{plain}
\pagenumbering{arabic}
\vspace*{-2.4cm}
\vspace{1.0cm}
\begin{center}
{\Large \bf 
Differential Equations for Scaling Relation in}
\end{center}
\vspace{-7.0mm}
\begin{center}
{\Large \bf $N=2$ Supersymmetric SU(2) Yang-Mills Theory }
\end{center} 
\vspace{-7.0mm}
\begin{center}
{\Large \bf Coupled with Massive Hypermultiplet }
\end{center} 
\lineskip .80em
\vskip 4em
\normalsize
\begin{center}
{\large Y\H uji Ohta}
\end{center}
\vskip 1.5em
\begin{center}
{\em Research Institute for Mathematical Sciences }
\end{center}
\vspace{-11.0mm}
\begin{center}
{\em Kyoto University}
\end{center}
\vspace{-11.0mm}
\begin{center}
{\em Sakyoku, Kyoto 606, Japan.}
\end{center}
\begin{abstract}
Differential equations for scaling relation of prepotential 
in $N=2$ supersymmetric SU(2) Yang-Mills theory coupled with 
massive matter hypermultiplet are proposed 
and are explicitly demonstrated in one flavour ($N_f =1$) theory. 
By applying Whitham dynamics, 
the first order derivative of the prepotential over the $T_0$ 
variable corresponding to the mass of the hypermultiplet, 
which has a line integral representation, 
is found to satisfy a differential equation. As the result, 
the closed form of this derivative can be obtained by solving this 
equation. In this way, 
the scaling relation of massive prepotential is established. 
Furthermore, as an application of another differential equation 
for the massive scaling relation, the 
massive prepotential in strong coupling region is derived. \\
PACS: 11.15.Tk, 11.15.Me, 12.60.Jv, 02.30.Hq.

\end{abstract}
\end{titlepage}


\begin{center}
\section{Introduction}
\end{center}

\renewcommand{\theequation}{1.\arabic{equation}}\setcounter{equation}{0}

It is well-known that the low energy effective action of 
$N=2$ supersymmetric Yang-Mills theory is described in terms of 
holomorphic prepotential $\F$.\cite{Sei} In this case, 
the perturbative part of the prepotential is not modified beyond 
one-loop order according to the non-renormalization 
theorem,\cite{GGRS,HST1,HST2,HSW} but 
is known to be affected by instantons. 
In the case of SU(2) gauge group, 
Seiberg and Witten \cite{SW1,SW2} proposed a general 
prescription to determine the non-perturbative prepotential 
with the aid of Riemann surface of genus one. Based on their 
observation, Klemm {\em et al.} \cite{KLT} determined the 
instanton corrected prepotential by adopting a method of 
Picard-Fuchs equation, which was often used in mirror symmetry 
of Calabi-Yau manifold. \cite{Yau,Can,Can2} 
On the other hand, Matone \cite{Mat} derived a 
recurrence formula of the instanton expansion coefficients of 
the prepotential by noticing a modularity. 
As a bonus, he obtained a quite simple relation between 
prepotential and moduli often referred as scaling relation. 
After this discovery, the existence of such relation in 
$N=2$ supersymmetric Yang-Mills theory coupled with or without 
massive quark hypermultiplets was pointed out independently by 
two groups.\cite{STY,EY} Sonnenschein {\em et al.} \cite{STY} proved 
it by noticing the homogeneity of the prepotential, while 
Eguchi and Yang \cite{EY} established the same result (\ref{hom}) 
in the language of Whitham dynamics.\cite{GKMMM,NT} 
However, in the framework of Whitham dynamics, 
we always encounter the problem of evaluating the derivative 
of $\F$ over the $T_0$ variable in order 
to establish a scaling relation when the mass of the hypermultiplets 
is not ignored. In the Whitham theory, this quantity is represented by 
a line integral interpolating two coverings of Riemann surface and 
therefore its evaluation is quite complicated, especially, 
in massive case. In fact, explicit calculation of this integral is not 
found in literatures. 

One of the aim of the paper is to give a solution to this problem. 
However, since the calculation in the case of a theory 
with $N_f$ massive flavours is very complicated even for SU(2), 
we treat only SU(2) $N_f =1$ case. 
In addition, extension to $N_f >1$ is straightforward, so the 
reader is recommended to try to proceed other cases. 
Our construction starts from comparing Wronskian of massive 
Picard-Fuchs equation with the modular invariant of 
Matone.\cite{Mat} These two are shown to be related 
by a Fuchsian differential equation. 
Applying Whitham theory of soliton to this equation, we can obtain a 
differential equation for $\pa \F/\pa T_0$. In this way, it is 
determined as a solution to this equation. 
The details are discussed in Sec. II. On the other hand, 
in the course of the calculus used in the derivation of 
this differential equation, we can find another simple differential 
equation, which indicates a relation between prepotential and moduli like that 
in five dimensions.\cite{KO} This equation is also a consequence of scaling relation 
in the massive theory. As an application, the prepotential in 
strong coupling region (dual prepotential) is derived in Sec. III. 
At first sight, 
the dual prepotential looks very complicated, but the massless limit 
coincides with that in the massless theory obtained by 
Ito and Yang.\cite{IY1} Sec. IV is a brief summary.

\begin{center}
\section{Scaling relation and Whitham hierarchy}
\end{center}

\renewcommand{\theequation}{2.\arabic{equation}}\setcounter{equation}{0}

\begin{center}
\subsection{The Picard-Fuchs equation}
\end{center}

First of all, let us recall the basics of the massive $N_f =1$ theory 
in Seiberg-Witten approach.\cite{SW2} In the case of SU(2), we can 
take two kinds of curves, one of which is elliptic type \cite{SW2} 
and the other is hyperelliptic type.\cite{HO} 
Though an elliptic curve is used in the next section, 
here we take the hyperelliptic curve. In this case, it is given by  
	\beq
	y^2 =(x^2 -u)^2 -\l^3 (x+m)
	,\lab{swmass}
	\eeq
where $u=\langle \mbox{tr }\phi^2\rangle$ is the moduli 
($\phi$ is the complex scalar field of $N=2$ chiral superfield), 
$m$ is the mass of the hypermultiplet and $\l$ 
is the mass scale parameter of this $N_f =1$ theory. 
For this curve, the Seiberg-Witten differential 1-form $\lambda$ 
is determined from the basic relation $d\lambda /du \propto dx/y$. 
Taking into account of the numerical normalization factor, we 
find\cite{HO,O}  
	\beq
	\lambda =\frac{\sqrt{2}}{4\pi i}\frac{xdx}{y}\left[
	\frac{x^2 -u}{2(x+m)}-2x\right]
	.\eeq
In general, 
Seiberg-Witten 1-form in a theory involving massive 
matter hypermultiplets in the fundamental representation 
of the gauge group is endowed with pole structure whose 
residue is linearly proportional to the masses of the 
hypermultiplets.\cite{SW2,HO} 

Seiberg-Witten ansatz requires that the vacuum expectation value of 
$\phi$ and its dual are quantum mechanically given by the two periods 
	\beq
	a=\oint_{\alpha}\lambda,\ a_D =\oint_{\beta}\lambda 
	\lab{twoperio}
	,\eeq
respectively, along the canonical basis ($\alpha \cap \beta =+1$) 
of 1-cycles on (\ref{swmass}). 
Then the periods satisfies the Picard-Fuchs equation 
	\beq
	\frac{d^3\Pi}{du^3}+X \frac{d^2\Pi}{du^2}+Y\frac{d\Pi}{du}
	=0
	,\lab{massivePF}
	\eeq
where 
	\beqa
	& &X=\frac{d}{du} \ln \frac{\Del}{4m^2 -3u},\nm\\
	& &Y=-\frac{8}{\Del}\left[4(2m^2 -3u)+3\frac{3m\l^3 -
	4u^2}{4m^2 -3u}\right]
	.\eeqa
Here, 
	\beq
	\Del (u)=256u^3 -256m^2 u^2 -288m\l^3 u +256m^3 \l^3 +27\l^6
	\lab{massivediscr}
	\eeq
is the discriminant of the curve (\ref{swmass}).

\begin{center}
\subsection{Differential equations for scaling relation}
\end{center}

Differential equation for Wronskian can be used to make a relation 
to prepotential. The reader might know the example in the 
context of mirror symmetry presented by Candelas 
{\em et al.},\cite{Can,Can2} who used ``Wronskian'' in order to 
make a contact with Yukawa coupling in complex structure moduli space. 
Similar presentation is possible also in $N=2$ supersymmetric Yang-Mills 
theory coupled with or without massless matter hypermultiplets. 
In the case of SU(2), integral of Wronskian of Picard-Fuchs equation 
yields the scaling relation of prepotential.\cite{EY} 
However, when hypermultiplets are massive, it is 
not easy to see such a simple relation connecting prepotential 
and moduli. We encountered in a similar problem 
in the five dimensional gauge theory.\cite{KO} 
Since we did not know a five dimensional analogue of the 
scaling relation in four dimensional gauge theory, we proposed a 
differential relation between prepotential and Wronskian. 
The method used in the course of this calculation reveals 
further aspects on scaling relation, 
provided it is applied to four dimensional gauge theory. 

For this, let us prepare the following two quantities
	\beq
	W=a^{\,'}a_{D}^{\,''}-a^{\,''}a_{D}^{\,'},\ 
	w=aa_{D}^{\,'}-a^{\,'}a_{D}
	,\lab{defwrons}
	\eeq
where $'=d/du$. The first equation is precisely the Wronskian 
for the third order Picard-Fuchs equation (\ref{massivePF}), 
while the integration of the second one produces the modular invariant 
of Matone\cite{Mat} 
	\beq
	\int_0 wdu =aa_D -2\F
	\lab{iiii}
	.\eeq

{\bf Remark:} {\em Modular invariance mentioned here is the sense of 
pure SU(2) theory.} \\ 
Here, the integral symbol indicates that it is 
an integration constant free integral, namely, the integration constant 
is set to zero. 
What we would like to clarify is a relation between (\ref{iiii}) and 
(integration of) $W$, so the first task is to try to connect $W$ and $w$. 

Fortunately, this is simply done by differentiating $w$ over $u$ repeatedly 
modulo the Picard-Fuchs equation (\ref{massivePF}), and 
we can find  
	\beq
	w^{\,''}+Xw^{\,'}+Yw=W
	\lab{wronskidif1}
	.\eeq 
It is interesting to notice that $w$ satisfies a Picard-Fuchs equation 
with $W$ as a source term. 
On the other hand, for $W$ we can easily obtain 
	\beq
	W^{\,'}+XW=0
	,\eeq
which implies  
	\beq
	W=c\frac{4m^2 -3u}{\Del}
	,\lab{source}
	\eeq
where $c$ is an integration constant. Note that the discriminant of 
the curve appears in the denominator and a similar relation was noticed 
in five dimensions.\cite{KO} In order to fix $c$, 
we may use massless or double scaling limit as a 
boundary condition. For example, in the massless limit, it is 
well-known that $w=i3/(4\pi)$,\cite{IY2} therefore, 
from (\ref{wronskidif1}) 
	\beq
	c=-i\frac{16}{\pi}
	\lab{ci}
	.\eeq
Of course, $c$ must be uniquely fixed, and the same value is obtained 
from double scaling limit. (\ref{source}) with (\ref{ci}) plays an 
important role in the next section. 

In this way, we arrive at 
	\beq
	w^{\,''}+Xw^{\,'}+Yw=-i\frac{16}{\pi}\frac{4m^2 -3u}{\Del}
	.\lab{wronskidif2}
	\eeq
Since the solution to this equation gives $w$ and must be related with 
(\ref{iiii}), $w$ obtained from (\ref{wronskidif2}) is 
expected to give a scaling relation in this massive theory \cite{STY} 
	\beq
	a\frac{\pa \F}{\pa a}-2\F =-
	m\frac{\pa \F}{\pa m}-\l \frac{\pa \F}{\pa \l}
	,\lab{massa}
	\eeq
where $\F$ is regarded here as a homogeneous function $\F =\F (a,\l ,m)$ 
in variables. Specifically, the solution of (\ref{wronskidif2}) will 
include informations on the right 
hand side of (\ref{massa}) and what we would like to do next is to 
extract such informations from (\ref{wronskidif2}). 
The best way to accomplish this is to introduce the Whitham 
theory.\cite{GKMMM,NT}

\begin{center}
\subsection{Relation to Whitham dynamics}
\end{center}

Let us consider a consequence of (\ref{wronskidif2}) in view of 
Whitham dynamics in $N=2$ Yang-Mills theory. For details of 
Whitham dynamics in the context of Yang-Mills theory, see Ref.14-16. 

Let $T_n$ $(n\in \mbox{\boldmath$N$}\cup\{0\})$ be time variables coupled 
to $(n+1)$-th order pole of Seiberg-Witten differential 1-form. 
In the SU(2) $N_f =1$ theory, the prepotential is available from 
the relation 
	\beq
	\frac{\pa \F}{\pa a}=\oint_{\beta}\lambda , \ 
	\frac{\pa \F}{\pa T_n}=-2\pi i \mbox{ res}\left(z^{-n}
	\lambda\right),\ \frac{\pa \F}{\pa T_0}=-2\pi i\int_{z_* =-m}^{
	z=-m}\lambda
	,\lab{dayon}
	\eeq
where residue is evaluated at $x=1/z =\infty$ and $z_* (=1/x)$ is the 
coordinate on the other sheet of the curve. At $x=\infty$, $\lambda$ is 
expanded as 
	\beq
	\lambda =\left[-\sum_{n>0}^{\infty}nT_n z^{-n-1}+T_0 z^{-1}
	-\frac{1}{2\pi i}\sum_{n>0}^{\infty}\frac{\pa \F}{\pa T_n}
	z^{n-1}\right]dz
	\lab{lam}
	.\eeq
Then, Whitham dynamics in $N=2$ Yang-Mills theory 
implies the homogeneity relation of the prepotential\cite{EY} 
	\beq
	a\frac{\pa \F}{\pa a}-2\F =-T_0 \frac{\pa \F}{\pa T_0}-
	T_1 \frac{\pa \F}{\pa T_1}
	\lab{hom}
	.\eeq
Note that the right hand side of (\ref{hom}) should be identified with 
that in (\ref{massa}). 

In the framework of Whitham hierarchy, 
$T_0$, $T_1$ and $\pa \F/\pa T_1$ are read from (\ref{lam}) as 
	\beq
	T_0 =-i\frac{m}{4\sqrt{2}\pi},\ T_1 =i\frac{3}{4\sqrt{2}\pi}
	,\ \frac{\pa \F}{\pa T_1} =\sqrt{2}\left(\frac{m^2}{4}-u
	\right)
	.\lab{egu}
	\eeq 
Due to our normalization, the numerical factors are different from 
those used by Eguchi and Yang.\cite{EY} 
Note that $T_0$ vanishes when $m=0$. 
This simplifies (\ref{hom}) and because of this, the scaling 
relation in the massless theory is easily determined by 
using Whitham hierarchy. In this case, it is not necessary to know 
$\pa \F /\pa T_0$, but in massive case it must be known. 
However, in a massive theory the calculation of $\pa \F/\pa T_0$ requires a 
care because the third equation in (\ref{dayon}) is a line 
integral from one covering of the Riemann surface to the other. 
In the case at hand, this integral 
consists of two pieces of integral from $-m$ to $\infty$ on one 
sheet and a copy of it on the other sheet (when the mass vanishes, it 
reduces to a familiar integral \cite{EY}). Due to the pole 
of $\lambda$ at $x=-m$, 
the evaluation of it is not easy and it is therefore still on 
challenging stage. However, we can develop another method and 
show that $\pa \F/\pa T_0$ satisfies a differential 
equation with the aid of (\ref{wronskidif2}). Then $\pa \F/\pa T_0$ 
is obtained as a solution to this equation. 

Differentiation of (the left hand side of) (\ref{hom}) 
gives $w$, so a differential equation 
for $\pa \F/\pa T_0$ can be obtained by substituting $w$ 
calculated from (\ref{hom}) with (\ref{egu}) into (\ref{wronskidif2})
	\beq
	\Theta^{\,'''}+X\Theta^{\, ''}+Y\Theta^{\,'}=
	\frac{4\sqrt{2}}{m}\left[i\frac{\pi}{2}W+\frac{6}{\Del}\left(
	-8m^2 +3\frac{3m\l^3 -4u^2}{4m^2 -3u}\right)\right]
	,\lab{wronskidif3}
	\eeq
where $\Theta =\pa\F /\pa T_0$. It is easy to get a solution to this 
equation 
	\beq
	\Theta^{\,'}=c_1 a^{\,'}+c_2 a_{D}^{\,'}+i\frac{2\sqrt{2}\pi}{m}w
	+i\frac{3\sqrt{2}\pi}{2m}\left[a_{D}^{\,'}\int_0 
	\frac{a^{\,'}Z}{4m^2 -3u}du-a^{\,'}\int_0 \frac{a_{D}^{\,'}Z}{
	4m^2 -3u}du\right]
	,\lab{soooo}
	\eeq
where $c_i$ are integration constants and 
	\beq
	Z=-8m^2 +3\frac{3m\l^3 -4u^2}{4m^2 -3u}
	.\eeq
In the derivation of (\ref{soooo}), we have used the fact that a second 
order differential equation 
	\beq
	\frac{d^2 y}{dx^2}+P(x)\frac{dy}{dx}+Q(x)y=R(x)
	\eeq 
with any function $P(x),Q(x)$ and $R(x)$ has a general solution 
in the form 
	\beq
	y=c_1 y_1 +c_2 y_2 -y_1 \int_0 \frac{y_2 R(x)}{W(y_1 ,y_2 )}dx
	+ y_2 \int_0 \frac{y_1 R(x)}{W(y_1 ,y_2 )}dx
	,\eeq
where $y_i$ are two independent solutions in the case of $R(x)=0$ and 
$W(y_1 ,y_2)$ is its Wronskian. Since (\ref{wronskidif3}) without the 
right hand side is nothing but the massive Picard-Fuchs equation, the above 
$y_i$ may be chosen as $a^{\,'}$ and $a_{D}^{\,'}$. 
Then $W(y_1 ,y_2)$ is identified 
with $W$ defined in (\ref{defwrons}). Furthermore, (\ref{source}) is 
used to arrive at the final expression (\ref{soooo}).   

Let us see the massless limit of (\ref{soooo}). In this limit, naively, 
the factor $1/m$ diverges, but we leave it 
for the moment. When $m$ vanishes, the integrals can be easily 
evaluated and the resulting terms cancels out the third term in 
(\ref{soooo}). Therefore, it follows that 
	\beq
	\left.\frac{\pa \F}{\pa T_0}\right|_{m\rightarrow 0}=
	c_1 a+c_2 a_D +\mbox{const}.
	\lab{824}
	\eeq
This result reflects the fact that (\ref{wronskidif3}) 
reduces to the total differentiation of the massless Picard-Fuchs equation 
because of vanishing of the right hand side of 
(\ref{wronskidif3}) for $m\rightarrow 0$. 
Actually, $\pa \F/\pa T_0$ corresponds to 
$\pa \F/\pa m$, so all constants in (\ref{824}) should be zero 
for $m\rightarrow 0$. 

On the other hand, integrating (\ref{soooo}) for $m\neq 0$, we obtain 
	\beq
	aa_D -2\F =i\frac{m}{6\sqrt{2}\pi}(c_1 a+c_2 a_D )+c_3 -\frac{1}{4}
	\int_0 \left[a_{D}^{\,'}\int_0 
	\frac{a^{\,'}Z}{4m^2 -3u}du-a^{\,'}\int_0 \frac{a_{D}^{\,'}Z}{
	4m^2 -3u}du\right]du 
	,\lab{korekore}
	\eeq
where $c_3$ is an integration constant, with the aid of 
(\ref{hom}) and (\ref{egu}). This is the general form of 
massive scaling relation. 

{\bf Remark:} {\em If the prepotential is known,\cite{O} $c_i$ are easily 
determined as 
	\beq
	c_1 =-3\pi in' ,\ c_2 =3\pi in,\ 
	c_3 =-\frac{m^2}{16}\left(i+4i\ln 2 -2\pi nn'\right)
	,\eeq
where $n,n' \in \mbox{\boldmath$Z$}$ are the winding numbers of 
1-cycles around the pole corresponding to $x=-m$ of $\lambda$. 
These constants supports the massless limit behaviour 
of $\pa \F/\pa T_0$. 
However, precisely, these integration constants must be determined 
from comparing it with the result of lower order expansion of 
the third equation in (\ref{dayon}).}

\begin{center}
\section{Massive prepotential in strong coupling regime}
\end{center}

\renewcommand{\theequation}{3.\arabic{equation}}\setcounter{equation}{0}

Next, let us study the prepotential in the strong coupling 
region following to the technology recently 
developed in the five dimensional gauge theory.\cite{KO} 
The basic tool in our case is (\ref{source}). 

For later convenience, we take the elliptic curve\cite{SW2}  
	\beq
	 y^2 =x^2 (x-u)+\frac{1}{4}m \l^3 x-
	\frac{\l^6}{64}
	,\lab{mascur}
	\eeq
whose discriminant coincides with (\ref{massivediscr}). 
The Seiberg-Witten differential 1-form is given by 
	\beq
	\lambda =\frac{\sqrt{2}}{8\pi}\frac{dx}{y}\left[2u-3x-\frac{m\l^3}{4x}
	\right]
	.\eeq
For this curve, we choose the 1-cycles on the surface (\ref{mascur}) 
as Ito-Yang cycles,\cite{IY1} which reduce for large $u$ with 
vanishing mass 
	\beq
	\alpha:\ -i\frac{\l^3}{8\sqrt{u}}\longrightarrow 
	 +i\frac{\l^3}{8\sqrt{u}},\ \beta:\ u\longrightarrow 
	-i\frac{\l^3}{8\sqrt{u}}
	.\eeq
Also in this case, the periods are defined by (\ref{twoperio}) and 
satisfy (\ref{massivePF}). 

To find strong coupling regime, let us decompose the discriminant as 
	\beq
	\Del (u)=256\prod_{i=1}^{3}(u-e_i)
	,\lab{eq}
	\eeq
where $e_i$ are given by the vanishing points of 
$\Del (u)=0$, which correspond to the strong coupling 
regime. $e_i$ can be easily obtained by solving 
$\Del (u)=0$ for $u$, but we should take care of the derivation. 
In general, any cubic equation in $x$ 
	\beq
	x^3 +ax^2+bx+c=0
	,\lab{cu}
	\eeq
where $a,b$ and $c$ are some constants independent of $x$, has 
three independent solutions. One of them is 
	\beq
	x=-\frac{2^{1/3}(-a^2 +3b)}{3
        \left[-2a^3 +9ab-27c+\sqrt{4(-a^2 +3b)^3 + 
        ( -2a^3 +9ab-27c )^2}\right]^{1/3}}+
	\cdots
	,\lab{formulaa}
	\eeq
where $\cdots$ means omission of remaining terms. 
However, if the denominator vanishes
	\beq
	-2a^3+9ab-27c+\sqrt{4\left(-a^2 +3b\right)^3 + 
      \left(-2a^3 +9ab-27c\right)^2}
	=0
	,\eeq
i.e., 
	\beq
	\left(a^2 - 3b \right)^3=0
	,\lab{koreda}
	\eeq
the formula (\ref{formulaa}) is not valid any more. In this case, one 
must start again from (\ref{cu}) under the condition (\ref{koreda}). 

In the case at hand, the condition (\ref{koreda}) for (\ref{massivediscr}) 
corresponds to 
	\beq
	m^3\left( 2m +3\l \right)^3
	\left( 4m^2 -6m\l+9\l^2 \right)^3 
	=0
	,\lab{aree}
	\eeq
which implies that the directly obtained location under $m\neq 0$ 
(c.f. (\ref{formulaa})), 
does not reflect precise massless limit of the massive discriminant. 
To understand more illustratively, let us set $u=x+iy$, where 
$x,y\in \mbox{\boldmath$R$}$. Then the equation $\Del (u)=0$ 
produces $3(x-m^2 /3)^2 -y^2 =m(8m^3 +27)/24$. 
For a non-zero $m$, this is nothing but hyperbolic curves, but for 
$m=0$ it becomes to crossing lines. Clearly, the zeros 
of the discriminant at $m=0$ behave singular because transition 
from hyperbolic curve to lines is always suppressed.  

In fact, though the correct massless location in the strong 
coupling regime must be 
	\beq
	e_{1}^{(0)}
	=-\frac{3\l^2}{2^{8/3}},\ 
	e_{2}^{(0)}=\frac{3(1-i\sqrt{3})\l^2}{2^{11/3}} ,\ 
	e_{3}^{(0)}=\frac{3(1+i\sqrt{3})\l^2}{2^{11/3}} 
	\lab{masslesspoint}
	,\eeq
these can not be obtained from (\ref{formulaa}) with $m =0$. 
Therefore, (\ref{aree}) distinguishes regions in quantum moduli space 
at $m =0$ and $m \neq 0$. 
To see a connection with massless theory, for example, 
we realize the zero locus of massive discriminant as a 
small mass perturbation in the form $e_i =e_{i}^{(0)}+(\mbox{series in }
m )$, where this series converges for $|m|<1$. 
Substituting this into the equation $\Del =0$ and 
equating coefficient of powers in $m$ to zero, one finds 
	\beqa
	& &e_1 =e_{1}^{(0)}-\frac{m}{2^{1/3}}\l +\frac{m^2}
	{3}-\frac{4}{27}\frac{2^{1/3}m^3}{\l}+\frac{
	4}{81}\frac{2^{2/3}m^4}{\l^2}+\cdots,\nm\\
	& &e_2 =e_{2}^{(0)}-\frac{(1+i\sqrt{3})}{2^{4/3}}m\l
	+\frac{m^2}{3} +\frac{2^{4/3}(1-i\sqrt{3})m^3}{27\l}+\cdots,\nm\\
	& &e_3 =e_{3}^{(0)}+\frac{(1-i\sqrt{3})}{2^{4/3}}m\l
	+\frac{m^2}{3}+\frac{2^{4/3}(1+i\sqrt{3})m^3}{27\l}+\cdots 
	.\lab{massper}
	\eeqa
As a final check, (\ref{eq}) must be satisfied, but this can be 
easily confirmed by order by order in $m$ greater than 
$m^3$. Note that (\ref{massper}) coincides with the massless 
strong coupling points for $m =0$ and thus 
the above $e_i$ are the expected ones for small but 
finite mass. Below, to make a contact with 
the result of Ito and Yang,\cite{IY1} $e_1$ is chosen as a representative of 
strong coupling regime. On the other hand, for $m$ greater than or 
equal to 1, (\ref{formulaa}) is available to derive zeros of the 
discriminant. 

{\bf Remark:} {\em In the case of SU(2) gauge group, 
this situation is characteristic to the $N_f =1$ theory. For $N_f =2$ and 
$3$ theories' discriminants, we do not encounter with such sensitive 
problem in the determination of strong coupling region. }

Performing differential calculation between periods and dual prepotential 
	\beqa
	& &\frac{da}{du}=\frac{\F_{D}^{\, ''}}{u^{\,'}},\nm\\
	& &\frac{d^2 a}{du^2}=\frac{1}{u^{\,'3}}\left(
	\F_{D}^{\,'''}u^{\,'}-\F_{D}^{\,''}u^{\,''}\right),\nm\\
	& &\frac{da^3}{du^3}=\frac{1}{u^{\,'5}}\left[
	\left(\F_{D}^{(4)}u^{\,'}-\F_{D}^{\,'''}u^{\,'''}\right)u^{\,'}-3
	u^{\,''}\left(\F_{D}^{\,'''}u^{\,'}-\F_{D}^{\, ''}u^{\, ''}
	\right)\right]
	,\eeqa
where $'=d/da_{D}$ and $\F_{D}$ is the prepotential defined by  
	\beq
	a=\frac{d\F_{D}}{da_{D}}
	\lab{dualprepoten}
	,\eeq
and using (\ref{massivePF}) for $\Pi =a$ with inverse relation $du/da_D$ 
(see also Ref.17), we can arrive at the differential equation for $\F_D$
	\beq
	\F_{D}^{(4)}+u^{\,'}\left(X-\frac{3u^{\, ''}}{u^{\, '2}}\right)
	\F_{D}^{\, '''}=0
	\lab{prere}
	.\eeq
This equation is integrated to give 
	\beq
	\F_{D}^{\, '''}=c\frac{4m^2 -3u}{\Del}u^{\, '3}
	\lab{sim}
	,\eeq
where $c$ is an integration constant to be determined later. The 
factor $(4m^2 -3u)/\Del$ corresponds to the Wronskian $W$ defined 
in (\ref{defwrons}). 

Since $\F_{D}^{\, '''}$ manifestly vanishes at $u=4m^2 /3$, $\F_{D}$ is 
represented by a linear combination of $a_{D}^2 ,a_D $ and $1$, 
although $O(a_{D})$-terms can be neglected, thus it follows 
immediately from (\ref{dualprepoten}) that $a \propto a_{D}$. 
This induces the trivial monodromy 
	\beq
	\left(\begin{array}{c}
	a \\	
	a_D 
	\end{array}\right) \longrightarrow \left(\begin{array}{cc}
	1 & 0\\
	0 &1 
	\end{array}\right)\left(\begin{array}{c}
	a \\	
	a_D 
	\end{array}\right) 
	,\eeq
which indicates that the indicial indices of the Picard-Fuchs 
equation at $u=4m^2 /3$ are integers and the BPS spectrum is unchanged at 
this point. 

{\bf Remark:} {\em The same monodromy can be seen from a 
version of (\ref{sim}) in weak coupling region.} 

As is easy to find that, only what we need is a 
function $u=u(a_D )$ to determine $\F_D$, but is obtainable from 
inverting the solution of the Picard-Fuchs equation. Therefore, 
substituting these data into (\ref{sim}) and 
triply integrate it, we will be able to determine $\F_{D}$. 
On the other hand, 
it would be sufficient to once determine $c$ at a representative point 
in the moduli space. Since $c$ has mass dimension zero, it can be 
regarded as a pure number. In fact, $c$ can be determined at the 
massless point and the result 
	\beq
	c=\frac{i}{16\pi}
	\eeq
follows from comparing (\ref{sim}) with the massless 
prepotential.\cite{IY2} 

The next one to be done is to calculate $u'$, but is a easy task. 
Since $u' =(da_D /du)^{-1}$, it is sufficient to obtain the 
solution to the Picard-Fuchs equation at $u=e_1$. With the help of 
(\ref{eq}), it is found by a linear combination 
	\beq
	\frac{da_D}{du}=\rho_1 \varphi_1 +\rho_2 \varphi_2
	\lab{rhoone}
	,\eeq
where	
	\beqa
	\varphi_1 &=&1-\frac{6u}{4m^2 -3e_1}-\frac{u^2}{e_1 
	(2e_1 -e_2 )(2e_1 -e_3 )(3e_1 -4m^2 )^2}\left[9e_{1}^2 
	(36e_1 -e_2 -e_3 )\right.\nm\\
	& &\left.-12(112e_{1}^2 -3e_1 e_2 -3e_1 e_3 
	+e_2 e_3 )m^2 +512(3e_1 -m^2 )m^4 
	+36m (3e_1 -4m^2 )\l^3\right]-\cdots,\nm\\
	\varphi_2 &=&u -\frac{3e_{1}^2 (4e_1 -e_2 -e_3)-4(8e_{1}^2 
	-3e_1 e_2 -3
	e_1 e_3 +e_2 e_3)m^2 }{2e_1 (2e_1 -e_2)(2e_1 -e_3)(3e_1 -4m^2 )}
	u^2 -\cdots 
	.\eeqa
Since these $\varphi_i$ are extremely complicated functions in 
the language of explicit $e_i$, i.e., right hand sides of $e_i$ in 
(\ref{massper}), we do not try to express $\varphi_i$ by using them. 
The constants $\rho_i$ are determined from asymptotic expansion of 
the period integral of $a_D$, i.e., 
	\beq
	\rho_1 = -\frac{1}{2^{1/6}\sqrt{3}\l},\ \rho_2 =
	-\frac{25}{9\l^3}\sqrt{\frac{2}{3}} 
	.\eeq
Integration produces $a_D$ itself with an integration constant, but 
from dimensional analysis, it is found that it has a unit mass 
dimension. Since all dependence of the scale parameter (instanton 
correction) should be entered in the right hand side 
of (\ref{rhoone}), the integration constant must be 
proportional to $m$. In addition, since the massless 
theory does not have this term, it can be 
regarded as a characteristic feature in massive theory. Actually, it is a 
residue contribution from the massive meromorphic 1-form, and this is the 
case. Thus, 
	\beq
	a_D =2\pi in\ \mbox{res }(\lambda )+\int_0 \left(\rho_1 
	\varphi_1 +\rho_2 \varphi_2\right)du
	,\lab{rho2}
	\eeq 
where $n \in \mbox{\boldmath$Z$}$ is the winding number 
of the $\beta$-cycle which 
loops around the pole of the massive meromorphic 1-form and the 
residue is evaluated at $x=0$. 
Denoting $\widetilde{a}_D =a_D -2\pi in\ \mbox{res }(\lambda)$ and 
repeatedly solving (\ref{rho2}), one can arrive at the inverse 
relation $u=u(a_D)$. 

In this way, the result 
	\beqa
	\F_{D}&=&\frac{1}{2}c_1 \widetilde{a}_{D}^2 +c_2 \widetilde{a}_{D}
	+c_3 \nm\\
	& &-i\frac{\l^2}{\pi}\frac{\widetilde{a}_{D}^2}{(e_1 -e_2 )
	(e_1 -e_3)}\left[\frac{3}{2^{17/3}}(3e_1 -4m^2 )
	(\ln\widetilde{a}_{D}^2 
	-3)+\frac{1}{2^{29/6}3^{2/3}(e_1 -e_2 )(e_1 -e_3 )}\right. \nm\\
	& &\left[125(e_1 -e_2)(e_1 -e_3)(3e_1 -4m^2)+9\cdot 2^{7/3}
	(e_2 +e_3 -2e_1 )m^2 \l^2 \right.\nm\\
	& &\left.\left.+27\cdot 2^{1/3}(6e_{1}^2 +4e_2 e_3 -5e_1 
	(e_2 +e_3))\l^2\right]
	\frac{\widetilde{a}_D}{\l} +\cdots \right]
	,\eeqa
where $c_i$ are integration constants, follows from expanding (\ref{sim}). 
$c_2 \widetilde{a}_D +c_3$ may be neglected because this terms 
does not change the effective coupling constant, but $c_1$ is non-trivial. 
Dimensional analysis shows that $c_1$ is a pure number. 
Therefore, it is sufficient to choose it as a constant such that 
$\F_D $ for $m \rightarrow 0$ coincides with that 
in the massless theory. 
As a check, one can see that this $\F_D$ reduces to that of 
massless theory for $m \rightarrow 0$, provided 
	\beq
	c_1 =-\frac{i}{4\pi}\ln \l^2
	.\eeq

The reader might already noticed that actually this prepotential was 
obtained irrelevantly from details of $e_1$, although 
it was calculated under the assumption of small mass to 
see a connection with the massless theory. 
For general $m$ in strong coupling regime, only the 
differences of the final form of the prepotential are explicit value of 
$e_i$, $c_i$ and $c$. For instance, for a large mass case, 
it is enough to 
simply expand (\ref{formulaa}) near $m =\infty$ to 
get the zeros of $\Del$ and replace $e_i$ by, e.g.,   
	\beq
	e_1 =m^2 +\frac{\l^3}{8m}+\cdots,\ 
	e_2 =\l^{3/2}\sqrt{m}-\frac{\l^3}{16m}+\cdots,\ 
	e_3 =-\l^{3/2}\sqrt{m}-\frac{\l^3}{16m}+\cdots
	.\lab{largemass}
	\eeq   

The period $a_D$ can be obtained from simply substituting 
(\ref{largemass}) into (\ref{rhoone}). In this way, the dual prepotential 
for a large mass is calculated, but it's form is not so 
attractive for us because it is again written by a 
complicated function in $e_i$. For this reason, it would not be 
necessary to write down the dual prepotential 
for $|m|\mbox{\underline{\underline{$>$}}}1$, and even for other 
$N_f$ cases. 

\begin{center}
\section{Summary}
\end{center}

\renewcommand{\theequation}{4.\arabic{equation}}\setcounter{equation}{0}

In this paper, it has been shown that the Wronskian of 
Picard-Fuchs equation in massive theory is related with 
(differentiation of) Matone's modular invariant through a differential 
equation. By comparing the solution to this equation with the 
Whitham dynamics, we have found a closed form of $\pa \F/\pa T_0$, 
which has not been evaluated so far. In this way, we have found a 
general form of the massive scaling relation. 
In the pure SU(2) theory, the scaling relation of prepotential 
is known to be also obtained as an anomalous superconformal 
Ward identity,\cite{HW} therefore, it would be a natural question to ask 
whether (\ref{korekore}) can be obtained as ``an anomalous 
superconformal Ward identity'' in this $N_f =1$ massive theory.  

On the other hand, we have also found that the differential equation for 
the Wronskian gives a differential relation between prepotential 
and moduli, which is quite reminiscent of the one presented 
in the five dimensional gauge theory \cite{KO} 
and the dual prepotential is calculated from this equation. 

More detailed study of a relation between Wronskian and modular invariant 
of Matone will provide us more useful 
informations on the massive prepotentials also in the future. 

\begin{center}
\section*{Acknowlegment}
\end{center}

The author acknowledges Prof. K. Ito (YITP) for discussions on 
miscellaneous aspects of $N=2$ Seiberg-Witten models. 

\begin{center}

\end{center}

\end{document}